\documentclass[10pt,aps,prb,superscriptaddress,
twocolumn,showpacs,floatfix,notitlepage,nofootinbib]{revtex4-2}
\usepackage{amsmath}
\usepackage{amsfonts}
\usepackage{amssymb}
\usepackage{hyperref}
\usepackage{graphicx}
\usepackage{color}
\usepackage[mathscr]{euscript}
\usepackage{bbm}
\usepackage{braket}
\usepackage{amsmath}
\usepackage{amssymb}
\usepackage{wrapfig}
\usepackage{tabularx}
\usepackage[utf8]{inputenc}
\usepackage{booktabs}
\usepackage[table]{xcolor} 
\usepackage{siunitx} 

\usepackage{array}
\usepackage{natbib}
\usepackage{rotating}
\usepackage[utf8]{inputenc}
\usepackage[T1]{fontenc}
\usepackage{soul}

\hypersetup{colorlinks=true,linkcolor=magenta
	,citecolor=magenta, urlcolor=magenta}

\begin{document}


\title{Dynamic Sculpting of Photon Statistics in a Raman-Coupled Cavity QED System}

\author{Amir Rahmani}
\affiliation{%
	Center for Quantum-Enabled Computing, Center for Theoretical Physics, Polish Academy of Sciences, Al. Lotnik\'{o}w 32/46, 02-668 Warsaw, Poland}%

\author{Alireza Dehghani}
\affiliation{Department of Physics, Sahand University of Technology, Tabriz, Iran\,
}%
\affiliation{Department of Physics, Payame Noor University, Tehran, Iran\,}

\begin{abstract}

Controlling photon statistics is pivotal to many applications in classical and quantum photonics. While steady-state photon statistics have been widely studied, dynamically manipulating these statistical properties in time remains elusive. Herein, we propose a scheme to actively control photon statistics between bunching and antibunching regimes through quantum interference. Our approach relies on a Raman-coupled cavity-atom system with parametric amplification, where time modulating the detuning of the intermediate state allows us to either deepen the photon blockade or enhance photon bunching. Simultaneous modulation of the cavity and intermediate-state detunings produces antibunching that cannot be achieved under static conditions or by modulating either detuning alone. Because the modulation is periodic, the antibunching reappears at regular time intervals, enabling time-gated photon antibunching under continuous-wave operation at a repetition rate determined by the modulation frequency. These results show a versatile mechanism based on Raman-assisted quantum interference for dynamically engineering photon correlations.

\end{abstract}

\maketitle
\section{Introduction}
Quantum interference lies at the core of many non-classical phenomena of light, occurs when indistinguishable paths coherently superpose~\cite{Ficek2005}. Prominent examples include single-photon self-interference~\cite{Braig2003}, the Hong–Ou–Mandel effect due to two-photon interference~\cite{PhysRevLett.59.2044}, multiphoton interferometry and entanglement~\cite{RevModPhys.84.777} and demonstrations of quantum supremacy~\cite{supermass}, among the others. In the realm of light–matter interaction, quantum interference leads to fascinating non-classical phenomena such as the Purcell effect~\cite{Lodahl2015}, electromagnetically induced transparency~\cite{Fleischhauer2005}, slow light~\cite{Khurgin2009}, and the unconventional photon blockade~\cite{PhysRevLett.121.043601,PhysRevLett.121.043602}, to name a few.

One way to characterize non-classical light is to study its photon statistics~\cite{Carmichael2008}, which describe the arrival dynamics of light—specifically whether photons travel in bunches or stream individually. Standard characterization relies on the normalized second-order correlation function, $g^{(2)}(\tau)$, where a zero-delay value of $g^{(2)}(0) < 1$ traditionally denotes antibunching and $g^{(2)}(0) > 1$ denotes bunching. A primary mechanism for generating such photon antibunching is photon blockade, which refers to the suppression of simultaneous multi-photon transmission due to nonlinear photon-photon interactions or quantum interference \cite{Imamoglu1997,Zub2020}. In conventional photon blockade (CPB), a strong anharmonicity in the energy-level ladder prevents the resonant excitation of a second photon \cite{Birnbaum2005}. By contrast, unconventional photon blockade (UPB) occurs in weakly nonlinear systems when different excitation pathways interfere destructively, leading to photon antibunching even with modest nonlinearities \cite{Liew2010, Bamba2011, Flayac2015}, although the blockade it produces is typically short lived in the delay time~\cite{Wang2026}. These phenomena have applications in quantum computing and communication \cite{Couteau2023SinglePhoton}, quantum sensing and metrology~\cite{Couteau2023Metrology}, and quantum imaging~\cite{Pearce2026,Gatti2004Ghost,Lemieux2025}.

Controlling quantum interference, such as through the second-order coherence function, is crucial for quantum information processing~\cite{Aspuru-Guzik2012,wang18}. For example, non-unitary optical elements can go beyond the conventional unitary interference and enable continuous control from bosonic bunching to fermionic antibunching ~\cite{Li2021}. Extending this framework to the dynamic regime may provide controlled, time-dependent sweeping through different quantum statistics leading to information processing and computing in real time. Despite its significance, only a few studies have focused on the dynamical control of quantum interference, using approaches such as phase-change materials~\cite{spb4-kgmq}, tailored coherent optical field~\cite{doi:10.1126/sciadv.abj1916,LeJeannic2022,Ghosh2019,PhysRevA.110.023718}, single-atom cavity QED systems acting as dynamic photon turnstiles within the Jaynes-Cummings (JC) framework~\cite{braak08}, voltage-programmable electro-optic device \cite{Bankwitz2026}, bi-tone Floquet driving~\cite{PhysRevLett.129.043601,Geng24},anda Floquet coupler~\cite{19kw-wmdw}.

One important but less-explored nonlinear interaction is the  Raman coupling between an atom and a single cavity mode~\cite{Phoenix90,ScullyZubairy}. In this regime, the atom is coupled to the cavity field via a far-off-resonant intermediate state \cite{ScullyZubairy}. After adiabatic elimination of the intermediate level, the effective interaction Hamiltonian takes the form $\hat{H}_{\rm int} \propto \hat{n} (\hat{\sigma} + \hat{\sigma}^\dagger)$, where $\hat{n}=\hat{a}^\dagger\hat{a}$ and $\hat\sigma$ is the atomic lowering operator~\cite{Law1997, Gerry1999}. This interaction does not change the photon number directly, as $[\hat{H}_{\rm int}, \hat{n}]=0$; however, it does not conserve the total excitation number because the atomic state can flip without a corresponding photon exchange. This dispersively mediated coupling leads to a photon-number-dependent phase accumulation on the atomic state. Physically, it means that the cavity field does not exchange photons with the atom; instead, the atomic state is driven at a Rabi frequency that depends on the number of cavity photons. 
This leads to a conditional phase accumulation; namely, if the cavity field is in a superposition of different Fock states, the atomic state acquires a number-dependent phase. Such a coupling is qualitatively different from the Jaynes--Cummings (JC) models \cite{Carmichael2015}, where each photon exchange flips the atomic state and changes the photon number by $\pm1$. The  Raman interaction thus offers a unique platform for quantum
control~\cite{PhysRevLett.133.116901}, with applications spanning quantum information processing~\cite{Imamoglu1997,ScullyZubairy}, quantum communication~\cite{Chen2017,Xiong2025}, and quantum sensing~\cite{Huang2025,Shi2022}. In particular, Ramanbased systems have been demonstrated for quantum teleportation~\cite{Chen2007}, quantum
memory~\cite{Xiong2025}, and quantum-enhanced interferometry~\cite{Huang2025}.

\begin{figure}[t]
\centering
\includegraphics[width=\linewidth]{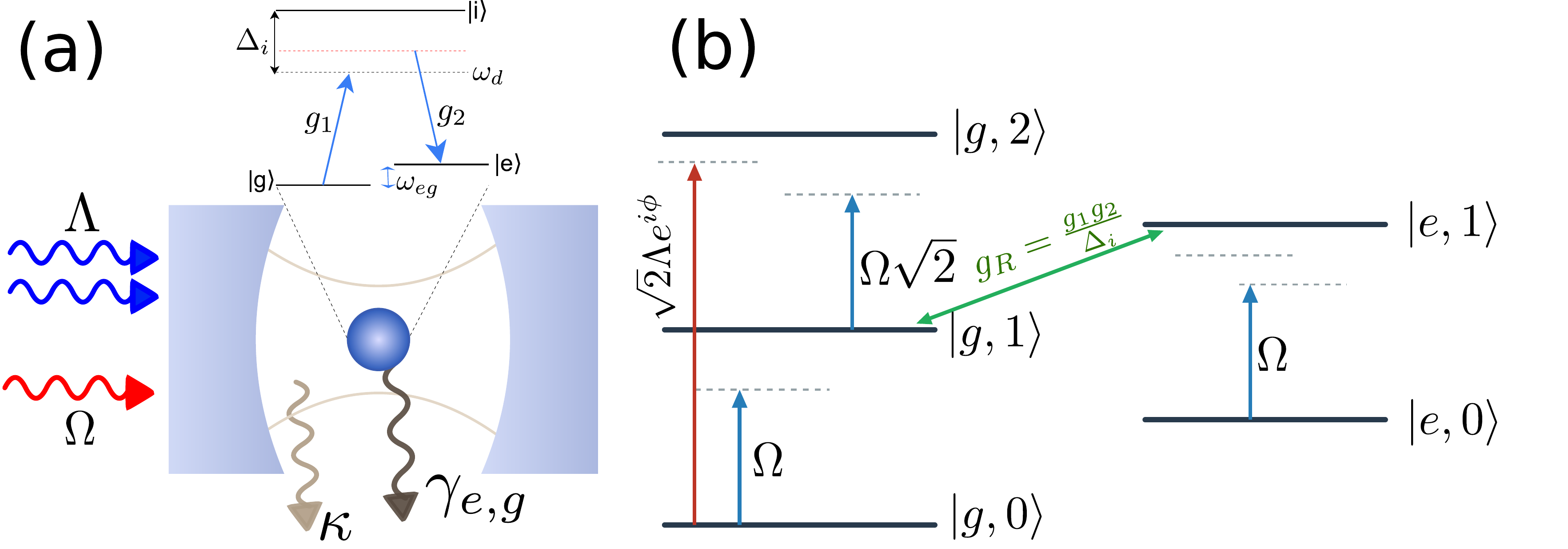}
\caption{(a) Schematic of the driven cavity-atom system under study. A three-level atom (grounds $|g\rangle$, $|e\rangle$, excited $|i\rangle$) is coupled to a single-mode cavity via a  Raman process of strength $g_{R}\propto g_1g_2$. The cavity is subject to a weak coherent drive (amplitude $\Omega$) and a degenerate parametric amplifier (strength $\Lambda$, phase $\phi$). Both the cavity and the atom undergo dissipation with rates $\kappa$ and $\gamma_{e,g}$, respectively. (b) Destructive quantum interference pathways for two-photon excitation. Pathway 1 (blue arrows): The coherent drive excites the system from $|g,0\rangle$ to $|g,1\rangle$ with amplitude $\Omega$, followed by a second drive-induced transition to $|g,2\rangle$ with amplitude $\Omega\sqrt{2}$. Pathway 2 (red arrows): The parametric drive directly couples $|g,0\rangle$ to $|g,2\rangle$ with amplitude $\Lambda e^{i\phi}$. The two pathways acquire relative phases that can be tuned via $\Lambda$ and $\phi$. 
}
\label{fig:new1}
\end{figure}

Here, we propose a scheme to dynamically control photon statistics via quantum interference by integrating a  Raman interaction into a driven, parametrically amplified cavity-QED system [Fig.~\ref{fig:new1}(a)]. Unlike previous works that focused only on parametric amplification dynamics~\cite{twhg-9tks,PhysRevA.96.053827,PhysRevA.100.023814}, we show that introducing Raman coupling provides a powerful mechanism for actively steering the emitted light from photon bunching to the antibunching regime by opening up new quantum interference paths. In the steady-state regime, increasing the Raman coupling degrades the antibunching and drives the system toward photon bunching, while a larger intermediate-state detuning restores it. However, time-dependent modulation of the detunings can overcome this limitation and restore antibunching. This effect originates from the atomic contribution to the system: the intermediate-state detuning enters the Hamiltonian through the Raman coupling and therefore has no effect when the atom is absent. Modulating this detuning changes the Raman interaction in time and, consequently, the interference responsible for the photon statistics. We further find that simultaneous modulation of the cavity and intermediate-state detunings produces antibunching that is not obtained under static conditions or by modulating either detuning alone.

We derive analytical conditions for achieving  photon antibunching through destructive interference and identify the key parameters governing the blockade. This is supplemented by solving the full Lindblad master equation, which completes the analysis for a detailed consideration of both the time-modulated and steady-state regimes. Although our theory relies on a Raman-coupled atom--cavity system, it can, in
principle, be extended to other active Raman modes, such as molecular
vibrations~\cite{Chattopadhyay2025} or Raman phonons~\cite{Forst2011}, provided
they can be mapped onto our effective model. The bi-tone modulation behaviour may also be possible in other driven nonlinear
cavity systems offering two or more independent modulation channels.

\section{THEORETICAL FRAMEWORK}
We consider a single-mode cavity containing a three-level atom, with the cavity mode parametrically pumped by a degenerate parametric amplifier (DPA) \cite{Milburn1981, DPA2010} and driven by a  coherent field. The total Hamiltonian (see the Supplementary Material for details) in the rotating frame at the drive frequency $\omega_d$ is
\begin{align}\label{eq:1}
\hat{H} &= \hbar \Delta_c \hat{a}^\dagger \hat{a} + \hbar \omega_{eg} \vert{}e\rangle\langle e\vert{} + \hbar \Delta_i \vert{}i\rangle\langle i\vert{}\nonumber\\ &\quad+ \hbar g_1 \left( \hat{a} \vert{}i\rangle\langle g\vert{} + \hat{a}^\dagger \vert{}g\rangle\langle i\vert{} \right) + \hbar g_2 \left( \hat{a} \vert{}i\rangle\langle e\vert{} + \hat{a}^\dagger \vert{}e\rangle\langle i\vert{} \right) \nonumber\\&\quad+ \hbar \Omega \left( \hat{a}^\dagger + \hat{a} \right) + \hbar \Lambda \left( \hat{a}^{\dagger 2} e^{-i\phi} + \hat{a}^2 e^{i\phi} \right)
\end{align}
%
where an atom with two ground states $\vert g\rangle$ and $\vert e\rangle$ and an excited state $\vert i\rangle$ is considered. The cavity mode $a$ couples the ground-state transitions $\vert g\rangle \leftrightarrow \vert i\rangle$ and $\vert e\rangle \leftrightarrow \vert i\rangle$ with coupling constants $g_1$ and $g_2$, respectively. We also define the detunings $\Delta_c = \omega_c - \omega_d$ and $\Delta_i = \omega_{ig} - \omega_d$.
Under large detuning $\Delta_i$, the intermediate state can be adiabatically eliminated, yielding a two-level effective model with coupling strength $g_R = g_1 g_2 / \Delta_i$:
\begin{align}
\hat{H}_{\mathrm{eff}}
&=
\hbar \Delta_c \hat{a}^\dagger \hat{a}
+
\hbar\omega_{eg}|e\rangle\langle e|
\nonumber\\&-
\hbar\hat{n}
\Big[
\frac{g_1^2}{\Delta_i}|g\rangle\langle g|
+
\frac{g_2^2}{\Delta_i}|e\rangle\langle e|
+
g_R\left(
|g\rangle\langle e|
+
|e\rangle\langle g|
\right)
\Big]\nonumber\\
&+\hbar \Omega \left( \hat{a}^\dagger + \hat{a} \right) + \hbar \Lambda \left( \hat{a}^{\dagger 2} e^{-i\phi} + \hat{a}^2 e^{i\phi} \right)\,,
\end{align}
The photon-number-dependent interaction $\hat{n}(|g\rangle\langle e|
+
|e\rangle\langle g|
)$ arises naturally from this elimination, as detailed in Supplementary Material. This derivation establishes the connection between the experimentally controllable parameters ($\omega_{eg}$, $g_{1,2}$, $\Delta_i$) and the effective coupling constants that appear in our theoretical model. The parametric drive (strength $\Lambda$, phase $\phi$) provides an additional two-photon pathway \cite{Yurke1989}, and the coherent drive $\hat{H}_d = \hbar\Omega(\hat{a}^\dagger+\hat{a})$ is applied directly to the cavity. For the analytical analysis, we consider the weak-driving regime (see below for details), which allows us to restrict our analysis to a small Hilbert-space manifold of the full Hamiltonian [Fig.~\ref{fig:new1}(b)]. Namely, we define the jump operators: $\hat{L}_g=\sqrt{\gamma_g}\vert{}g\rangle\langle i\vert{}$, $\hat{L}_e=\sqrt{\gamma_e}\vert{}e\rangle\langle i\vert{}$, and $\hat{L}_a=\sqrt{\kappa}\hat{a}$, which can be used to introduce an effective non-Hermitian Hamiltonian $\hat{H}_{\text{Non-Herm}}=\hat{H}-\frac{i}{2}\sum_{j \in \{g,e,a\}}\hat{L}_j^\dagger \hat{L}_j$, where the cavity and atomic decay rates are $\kappa$ and $\gamma_{e,g}$, respectively. For the numerical analysis, we solve the full Lindblad master equation using realistic parameters typical of cavity QED systems~\cite{Blais2004, Wallraff2004}:
\begin{align}\label{eq:fullmaster}
\frac{d\rho}{dt}
=&
-i[\hat{H},\rho]
\nonumber\\
&+
\sum_{j=\kappa,\gamma_g,\gamma_e}
\left(
\hat{L}_j\rho \hat{L}_j^\dagger
-\frac{1}{2}\hat{L}_j^\dagger\hat{L}_j\rho
-\frac{1}{2}\rho\hat{L}_j^\dagger\hat{L}_j
\right).
\end{align}
In the following two sections, we study the second-order correlation function defined as:
\begin{equation}\label{eq:jdsbued8}
g^{(2)}(t+\tau, t) = \frac{\langle \hat{a}^\dagger(t) \hat{a}^\dagger(t+\tau) \hat{a}(t+\tau) \hat{a}(t) \rangle}{\langle \hat{a}^\dagger(t) \hat{a}(t) \rangle \langle \hat{a}^\dagger(t+\tau) \hat{a}(t+\tau) \rangle}.
\end{equation}
This gives the correlation between  photon emissions at times $t$ and $t+\tau$. We consider zero time delay $\tau=0$ (equal time) across two regimes (we analyze the case of $g^{(2)}(\tau)$ in the Supplementary Material): the dynamical equal time correlation function $g^{(2)}(t) \equiv g^{(2)}(t,t)$ and its long-time steady-state limit $g^{(2)}(0) \equiv \lim_{t \to \infty} g^{(2)}(t,t)$. The numerical Results in the following sections are obtained from the full three-level Lindblad master equation and an effective Raman description is used for interpreting the dynamics.
\section{Steady-State Photon Statistics}
In this section, we present a detailed analytical and numerical study of photon statistics in our  Raman-coupled cavity-atom system with parametric amplification. Assuming a time-independent regime, we first derive the optimal conditions for destructive quantum interference to suppress two-photon excitation. This approach uses a two-photon truncation of the Hilbert space alongside an effective non-Hermitian Hamiltonian to account for cavity and atomic dissipations. This yields closed-form expressions for the optimal parametric gain $\Lambda$ and phase $\phi$ that lead to $g^{(2)}(0)\to 0$. We examine how Raman coupling enables control over quantum interference through the interaction between photonic and matter components. We then validate our analytical results against full numerical solutions of the Lindblad master equation. In particular, we explore the roles detuning $\Delta_i$ and $\Delta_c$, two-photon drive $\Lambda$ in shaping the photon blockade and bunching. 
The numerical simulations also allow us to map out the optimal parameter regions for strong antibunching, including the dependence on the parametric phase $\phi$ and gain $\Lambda$, as shown in Fig.~\ref{fig:new2}.

\subsection{Analytical Results}
The optimal conditions for parametric-drive-induced photon blockade have been established for a two-level atom in a cavity, where the cavity-atom coupling is of the Jaynes--Cummings form [48]. Here we derive the corresponding condition for the number-conserving Raman coupling, which introduces the intermediate-state detuning $\Delta_i$as an additional parameter. This detuning is the control channel we modulate in Sec.~IV, and the static analysis below establishes how it shapes the interference condition.

We work under weak driving conditions $\Omega,\Lambda\ll \kappa$. It allows to truncate the Hilbert space to at most two photons. Such low-excitation manifold is shown in Fig.~\ref{fig:new1}(b). The truncated basis $\left\{ |g,0\rangle,\; |g,1\rangle,\; |e,0\rangle,\; |e,1\rangle,\; |g,2\rangle \right\}$ captures all relevant pathways leading to two-photon excitation and the resulting quantum interference responsible for photon statistics. The wavefunction is
\begin{eqnarray}
|\psi\rangle = \sum_{n=0}^2 c_{g,n}|g,n\rangle + \sum_{n=0}^1 c_{e,n}|e,n\rangle.
\end{eqnarray}
This wavefunction ansatz has been extensively used in studies of unconventional
photon blockade in cavity-QED systems~\cite{Bamba2011,Carmichael2015}. The validity of this
truncation against full numerical solutions of the
Lindblad master equation is  shown in  Fig.~\ref{fig:new2}(a). The dynamical equations for the probability amplitudes are derived from the Schr\"odinger equation with an effective non-Hermitian Hamiltonian that includes decay, namely $i \partial_t |\psi\rangle=H_{\text{Non-Herm}}|\psi\rangle$.
For the cavity-driven case, and assuming $\Delta_i$ is very large, we obtain a set of steady state solutions (see the Supplemntary Material for details):
\begin{subequations}
	\begin{align}
	c_{g,1} = \frac{\Omega (\Lambda e^{-i\phi} - \delta_1)}{\delta_1 \delta_g - \Omega^2}\label{eq:6a}\,,\\
	c_{g,2} = \frac{\Omega^2 - \Lambda \delta_g e^{-i\phi}}{\sqrt{2}(\delta_1 \delta_g - \Omega^2)}\label{eq:6b}\,,
	\end{align}
\end{subequations}
$c_{e,1} = \frac{g_R}{\delta_e} c_{g,1}$ and $\quad c_{e,0} = -\frac{\Omega}{\omega_{eg}} c_{g,1}$. Here we define $\delta_e = \delta_2 - \frac{\Omega^2}{\omega_{eg}}$ and $\delta_g = \delta_1 - \frac{g_R^2}{\delta_e}$, $\delta_1 = \tilde{\Delta}_c - \frac{g_1^2}{\tilde{\Delta}_i}$, $\delta_2 = \tilde{\Delta}_c + \omega_{eg} - \frac{g_2^2}{\tilde{\Delta}_i}$,  $\tilde{\Delta}_c = \Delta_c - i\frac{\kappa}{2}$, and $\tilde{\Delta}_i = \Delta_i - i\frac{\gamma_g+\gamma_e}{2}$. We also assume  the weak-excitation assumptions $|c_{g0}| \gg |c_{g1}|$, $|c_{e0}| \gg |c_{g2}|, |c_{e1}|$, and $c_{g0}\approx 1$.
The second-order correlation function is $g^{(2)}(0) \approx \frac{2|c_{g,2}|^2 }{\left( |c_{g,1}|^2|\right)^2}$. The regime of particular interest is the photon blockade which requires $g^{(2)}(0)\to 0$. This  corresponds to destructive quantum interference between  two excitation pathways leading to the two-photon state. As shown in Fig.~\ref{fig:new1} (b), the first path is driven by the coherent field which excites the system from $|g,0\rangle$ to $|g,1\rangle$ with amplitude $\Omega$, followed by a second drive-induced transition to $|g,2\rangle$ with amplitude $\Omega\sqrt{2}$. The second pathway is induced by the parametric drive which directly couples $|g,0\rangle$ to $|g,2\rangle$ with amplitude $\sqrt{2}\Lambda e^{i\phi}$. The two pathways acquire relative phases that can be tuned via $\Lambda$ and $\phi$. When these amplitudes interfere destructively, the two-photon population $c_{g,2}$ vanishes, leading to photon blockade ($g^{(2)}(0)\to 0$). The condition is given by:
\begin{equation}\label{eq:cent}
\Lambda e^{-i\phi} = \frac{\Omega^2 \left[ \Omega^2 - \omega_{eg}(\tilde{\Delta}_c + \omega_{eg} - \frac{g_2^2}{\tilde{\Delta}_i}) \right]}{(\tilde{\Delta}_c - \frac{g_1^2}{\tilde{\Delta}_i})\left[ \Omega^2 - \omega_{eg}(\tilde{\Delta}_c + \omega_{eg} - \frac{g_2^2}{\tilde{\Delta}_i}) \right] + g_R^2 \omega_{eg}}\,.
\end{equation}
The explicit analytical expressions for the optimal $\Lambda$ and $\phi$ are derived and discussed in detail in the Supplementary Material.

\subsection{Numerical Results}
 
\begin{figure}[t]
\centering
\includegraphics[width=1\linewidth]{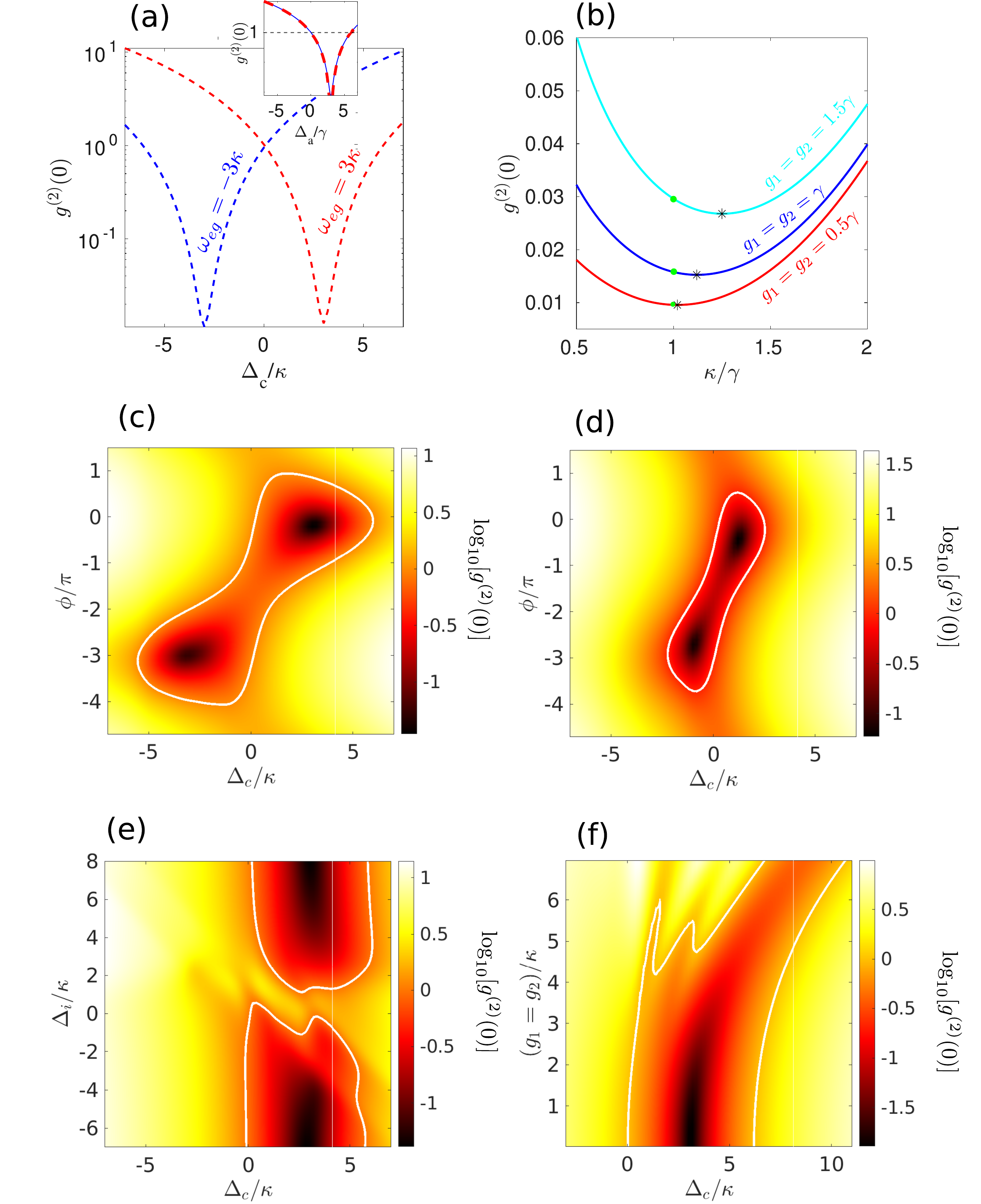}
\caption{(a): Second-order correlation function $g^{(2)}(0)$ versus detuning $\Delta_c/\kappa$ for two  different detuning $\omega{eg}$. We assume  $\Lambda$ and $\phi$ set to the condition in Eq.~(\ref{eq:cent}) for $\Delta_c=\omega_{eg}=\pm 3$. Other parameters are $\Omega=0.1\kappa$,  $g_1=g_2=\kappa$, $\gamma_{e}=\gamma_g=0.4\kappa$, $\Delta_i=20\kappa$, $\omega_{eg}=3\kappa$. The inset shows the agreement between the numerical (blue-solid line) and analytical (red-dashed line) results for weak drives. (b): The variation of $g^{(2)}(0)$ versus cavity decay rate. The parametric drive is set from Eq.~(7) with $\kappa=\gamma$ (the green dots). The star shows the the minima. (c): Density plot of $\log_{10}[g^{(2)}(0)]$ versus $\Delta_c/\kappa$ and $\phi/\pi$ for fixed $\Lambda=0.0032\kappa$. Other parameters are the same as panel a. (d): Same as (b) but with $\Lambda=0.008\kappa$. (e): Density plot of $\log_{10}[g^{(2)}(0)]$ versus $\Delta_c/\kappa$ and $\Delta_i/\kappa$. We assume $\phi=-0.1$ and other parameters are the same as in panel (b). (f):  Density plot of $\log_{10}[g^{(2)}(0)]$ versus $\Delta_c/\kappa$ and $(g_1=g_2)/\kappa$. Using parameters:$\Delta_i=20\kappa$, $\omega_{eg}=3\kappa$, $\gamma_e=\gamma_g=0.4\kappa$. In panels (c), (d), (e) and (f), the white curve shows the contour corresponding to $g^{(2)}(0)=1$.
}
\label{fig:new2}
\end{figure}

We now turn to the full numerical treatment of the system and analyze the photon statistics by solving the master equation (3). Figure~\ref{fig:new2} presents six panels (a)-(f) that systematically map the behaviour of \(g^{(2)}(0)\) in the Raman-coupled system.

Panel (a) shows \(g^{(2)}(0)\) versus the normalized cavity detuning \(\Delta_c/\kappa\) for two different values of the atomic detuning $\omega_{eg}=\pm3\kappa$ . Here, we assume a large $\Delta_i$, so the atomic excited state is numerically eliminated and the parameters \(\Lambda\) and \(\phi\) are set to the optimal condition (7). Two pronounced antibunching dips appear at $\Delta_c/\kappa = \pm 3$. Moving away from these values weakens the photon blockade, causing the light to become eventually a bunched state ($g^{(2)}(0) > 1$). The inset confirms excellent agreement between the analytical two-level truncation (red dashed) and the full Lindblad solution (blue solid) in the weak-drive regime. In panel (b) we fix the parametric drive and vary the cavity decay rate. For each coupling strength, $\Lambda$ and $\varphi$ are taken from Eq. (7) at $\kappa_0=\gamma$ and then held constant. Every curve passes through a minimum and rises on both sides, so the blockade degrades gradually rather than abruptly when the cavity linewidth differs from its design value. For $g_1 = g_2 = 0.5\gamma$ the correlation function stays around $0.01$ which comes from Eq. (7) itself. The decay rate enters through $\tilde{\Delta}_c = \Delta_c - i\frac{\kappa}{2}$ and hence through $\delta_g$, so changing $\kappa$ alters both the magnitude and the phase of the parametric drive that the interference condition demands. The minimum sits at $\kappa_0$ for the weakest coupling, but moves away as we set $g_1=g_2=\gamma$ and $g_1=g_2=1.5\gamma$. This probably reflects the approximations behind Eq. (7), but the relevant point here is that $\kappa$ provides an additional control knob, alongside $\Lambda$ and $\varphi$, for reaching the interference condition.

%
\begin{figure*}[t]
	\centering
	\includegraphics[width=1\linewidth]{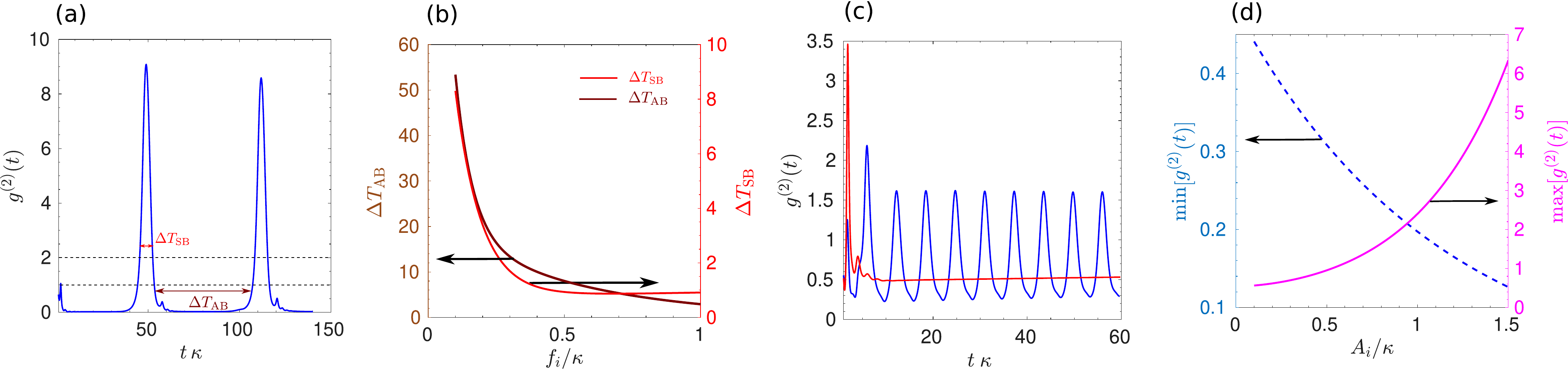}
	\caption{Time evolution and parametric dependence of the second-order correlation function. (a): $g^{(2)}(t)$ versus normalized time $t\kappa$ for sinusoidal modulation of $\Delta_i$. Two time windows are indicated; $\Delta T_{\text{AB}} (\Delta T_{\text{SB}})$ show the time window that light is antibunched (super-bunched))in a single repetition. We assume $\Delta_i$ is modulated as $\Delta_i+A_i \sin(f_i t)$. Using parameters are: $\Delta_c=\omega_{eg}=3\kappa$, $g_1=g_2=\kappa$, $\Omega=0.1\kappa$, $\Lambda=0.0035\kappa$, $\phi=-0.1$, $\gamma_e=\gamma_g=0.4\kappa$, $A_i=\Delta_i=6.5\kappa$ and $f_i=0.1/\kappa$. (b): The two time windows $\Delta T_{\text{AB}}$ and $\Delta T_{\text{SB}}$ vary exponentially with the modulation frequency $f_i$. (c): Periodic modulation of $g^{(2)}(t)$ using a different set of parameters. The steady-state result is shown in red for reference. Using parameters: $\Delta_i=2.2\kappa$, $\Delta_c=\omega_{eg}=2.6\kappa$, $g_1=g_2=\kappa$, $\Omega=0.1\kappa$, $\gamma_e=\gamma_g=0.4\kappa$, $\phi=-0.1$, $\Lambda=0.0045\kappa$, $A_i=0.8\kappa$ and $f_i\kappa=1$. 
	 (d): Minimum and maximum values of $g^{(2)}(t)$ within a single period as a function of the modulation amplitude $A_i$ in panel (c).}
	\label{fig:new3}
\end{figure*}

Panel (c) presents a density plot of \(\log_{10}[g^{(2)}(0)]\) in the two-dimensional space \((\Delta_c/\kappa,\ \phi/\pi)\) for a fixed parametric gain \(\Lambda = 0.0032\kappa\). The white contour marks \(g^{(2)}(0)=1\). A narrow region of strong antibunching (\(\log_{10}[g^{(2)}(0)] < -1\)) appears around \(\Delta_c/\kappa \simeq \pm 3\), with the optimal phase depending on the detuning. The width of the antibunching window is phase-sensitive, confirming that proper phase matching is essential for destructive interference.

Panel (d) shows the same density plot but with a larger parametric gain \(\Lambda = 0.008\kappa\). Compared to panel (c), the antibunching region shrinks and its minimum value increases (blockade weakens). This demonstrates that although \(\Lambda\) is optimal for a specific detuning, deviations from that detuning while keeping \(\Lambda\) fixed rapidly destroy the interference condition. The white contour encloses a smaller area, indicating higher sensitivity to parameter fluctuations.


Panel (e) displays $\log_{10}[g^{(2)}(0)]$ versus $\Delta_c/\kappa$ and $\Delta_i/\kappa$ (the intermediate-state detuning).  the antibunching region is concentrated at large $\Delta_i$, whereas the light becomes bunched around zero detuning ($\Delta_i \approx 0$). This reveals the usefulness of tuning detunings to control the statistical regime of the light. We also notice to an asymmetry with respect to the sign of the detunings which originates from the nature of the Raman coupling, which breaks the symmetry under a single sign flip. Only simultaneous reversal of both $\Delta_c$ and $\Delta_i$ restores the symmetry—a unique feature of this system. 

Finally, panel (f) shows \(\log_{10}[g^{(2)}(0)]\) versus \(\Delta_c/\kappa\) and \((g_1=g_2)/\kappa\) (which controls both the Raman coupling strength and the effective detunings). This panel directly visualises the detrimental effect of increasing the Raman coupling: the antibunching region is gradually eroded as the coupling becomes stronger, confirming that the static Raman interaction disrupts the destructive interference. The white contour \(g^{(2)}(0)=1\) shifts to lower detunings and eventually disappears for large coupling. To understand the underlying mechanism behind Raman coupling, we examine the role of the Raman coupling $g_R$, which enters to the condition (\ref{eq:cent}) through $\delta_g=\delta_1-\frac{g_R^2}{\delta_e}$. This contains the dispersive shift ($g_1/\Delta_i$), and increasing $g_{1,2}$ displaces $\delta_g$ whereby the condition (\ref{eq:cent}) becomes weak. This generates an enhancement in $g^{(2)}(0)$, signaling a regime of strong photon superbunching. This is further confirmed by the systematic suppression of antibunching with increasing $g_R$ observed in Fig.~\ref{fig:new2}f, where the Raman coupling degrades the destructive interference.

Taken together, the results of Fig.~\ref{fig:new2} establish the static "phase diagram" of the system. They reveal that the  intermediate-state detuning and Raman coupling act as a switch: tuning them allows one to access different regimes of photon statistics, either degrading or enhancing the photon blockade. Also,  there is a pronounced parameter asymmetry; namely, the lack of invariance under single-detuning sign flips provides a new knob for directional control of photon statistics, analogous to non-reciprocal effects in other quantum optical systems \cite{Dong2026}. These static results are essential for understanding the time-modulated regime presented in the next section.

\section{Dynamical photon statistics}
%

In this section, we extend our analysis to the time-dependent regime, where the intermediate-state detuning $\Delta_i$ and cavity detuning $\Delta_c$ are periodically modulated in time. We assume $\Delta_{i;c}(t) = \Delta^{0}_{i;c} + A_{i;c} \sin(f_{i;c}t)$, where $A_{i;c}$ and $f_{i;c}$ are the modulation amplitude and frequency, respectively. The physical origin of the time-dependent detuning lies in the ability to dynamically control the energy-level spacing of the atomic system through external fields---for instance, via ac-Stark shifts induced by control lasers, or through modulation of the cavity resonance frequency using electro-optic or optomechanical effects \cite{PhysRevA.110.023718,PhysRevLett.129.043601,Li2021}. This temporal modulation of $\Delta_i(t)$ directly affects the atom–cavity interaction, thereby providing a time-varying quantum interference condition that can dynamically sculpt the photon statistics.

Panel (a) of Fig.~\ref{fig:new3} shows the time evolution of $g^{(2)}(t)$ for a sinusoidally modulated $\Delta_i(t)$  with modulation amplitude $A_i = 6.5\kappa$ and frequency $f_i = 0.1\kappa$. Here, we select a set of parameters for which the steady-state regime lies deep in the antibunching region ($g^{(2)} < 1$). Under modulation, the dynamic quantum interference creates distinct temporal windows for photon statistics (it is worth noting that this window refers to the emission time $t$ and not to the delay $\tau$, the delay dependence of the correlation function is analysed in the Supplementary material). Interestingly, over the broad interval $\Delta T_{\text{AB}}$, destructive interference suppresses two-photon states, maintaining strong single-photon antibunching. We notice that the system sustains a remarkably long antibunching window that spans dozens of cavity lifetimes ($\Delta T_{\text{AB}} \gg 1/\kappa$). This cancellation momentarily can be broken down, producing sharp superbunching spikes up to $g^{(2)}(t) \approx 9$ within the narrow window $\Delta T_{\text{SB}}$. This dynamical tuning of the interference condition is analogous to the concept of ``coherent control'' in quantum optics, where time-dependent fields are used to manipulate quantum interference effects \cite{PhysRevA.100.023814}. 

Panel (b) displays the durations $\Delta T_{\mathrm{AB}}$ and $\Delta T_{\mathrm{SB}}$ as functions of the modulation frequency $f_i$. Both windows decrease with increasing $f_i$, following a power law form $\Delta T_{\mathrm{AB,SB}} \propto f_i^{-\alpha}$ with $\alpha \approx 1.2$ which the fitted exponent is determined from the numerical data. This behavior is consistent with the response of a driven dissipative system, where faster modulation leaves less time for the system to adiabatically follow the changing interference condition \cite{Lemonde2016}. The exponential decay reflects the finite response time set by the cavity and atomic decay rates $\kappa$ and $\gamma$, which establish a characteristic timescale $\tau_{\mathrm{resp}} \sim 1/\kappa$ for the system to reach its steady state. For modulation frequencies $f_i \gg \kappa$, the system cannot adiabatically track the instantaneous interference condition, and the amplitude of the $g^{(2)}(t)$ oscillations is suppressed. This provides a practical guide for choosing modulation parameters: to maximize the duration of the antibunching windows, one should operate at low modulation frequencies ($f_i \lesssim \kappa$), while higher frequencies can be used to generate rapid switching between statistical regimes at the cost of reduced window duration.
\begin{figure}[t]
	\centering
	\includegraphics[width=1\linewidth]{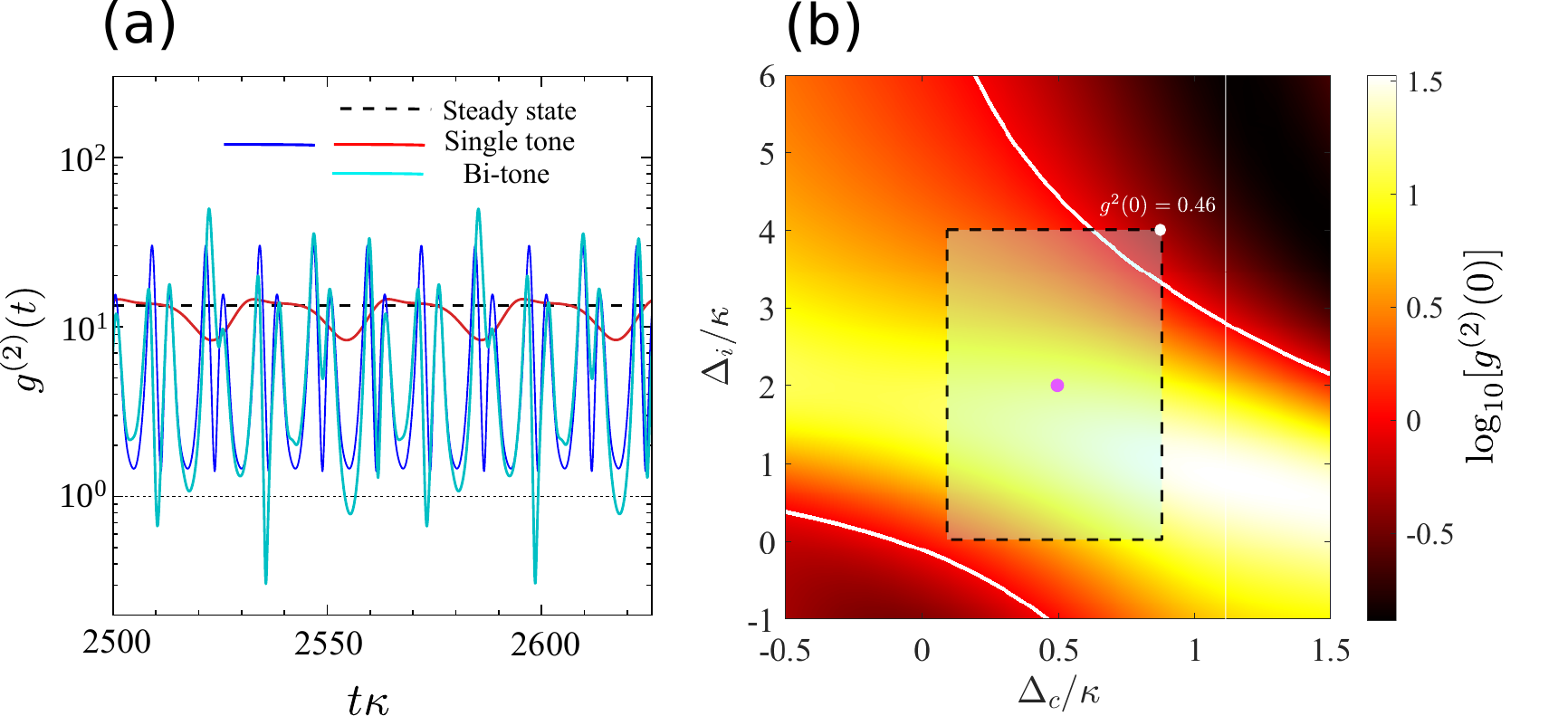}
	\caption{(a): Long-time (Floquet) regime of $g^{(2)}(t)$ for single tone and bi-tone modulation of $\Delta_i$ and $\Delta_c$. The steady-state regime (shown by the dashed line) is in the superbunching regime ($g^{(2)}(0)=13.4$). We assume $\Delta_i(t)=2+2\sin(0.5t)$ (the associated $g^{(2)}(t)$ shown in red) and $\Delta_c(t)=0.5+0.4\sin(0.2t)$ (the associated $g^{(2)}(t)$ shown in blue). The cyan curve shows $g^{(2)}(t)$ when both $\Delta_i$ and $\Delta_c$ are modulated. (b) The density plot of $g^{(2)}(0)$ in terms of $\Delta_i/\kappa$ and $\Delta_c/\kappa$. The magenta point shows the center of modulation: $(\Delta_c=0.5,\Delta_i=2)\kappa$. The rectangle shows the sweeping area where $g^{(2)}(t)$ sweeps in the bi-tone modulation. The white point at the upper-right corner shows the minimum value of $g^{(2)}(0)$ within the rectangle. Using parameters: $
		g_1=g_2=1.5\kappa,\quad
		\Omega=0.05\kappa,\quad
		\omega_{eg}=3\kappa,\quad
		\gamma_g=\gamma_e=0.4\kappa,\quad
		\Lambda=2.27\times10^{-3}\kappa,\quad
		\phi=-0.926.
		$}
	\label{fig:new4}
\end{figure}

Panel (c) shows how time-dependent detuning modulation overcomes static blockade limits. While the unmodulated steady state (red) yields only moderate antibunching ($g^{(2)} \approx 0.5$), periodically driving the intermediate-state detuning causes $g^{(2)}(t)$ to oscillate about the static value, with the minima dropping well below the unmodulated steady state. The modulation thus provides temporal access to deeper antibunching without retuning the operating point.

Panel (d) plots the minimum and maximum values of $g^{(2)}(t)$ within a single modulation period as a function of the modulation amplitude $A_i$ as we $\Delta_i$ is modulated. As $A_i$ increases, the system explores a wider range of effective coupling, leading to larger excursions of $g^{(2)}(t)$. The minimum value of $g^{(2)}(t)$ decreases monotonically with increasing $A_i$, reaching a minimum of $\approx 0.2$ at $A_i = 1.5\kappa$, while the maximum value increases correspondingly. The asymmetry between the minima and maxima reflects the nonlinear dependence of the interference conditions, which in turn depends on $\Delta_i(t)$. 
The monotonic decrease of $g^{(2)}_{\mathrm{min}}$ with $A_i$ suggests that larger modulation amplitudes are beneficial for achieving deeper transient antibunching, consistent with the idea that a wider sweep of Ramman coupling increases the probability of hitting the optimal interference condition.

Simultaneously modulating both $\Delta_c$ and $\Delta_i$ introduces a distinct
regime for the dynamics of the second-order correlation function. An example is
shown in Fig.~\ref{fig:new4}. Here the steady-state $g^{(2)}(0)$ lies in the
superbunching regime, indicated by the dashed line in panel (a). This corresponds
to the magenta point in panel (b), which shows the density map of $g^{(2)}(0)$ as
a function of $\Delta_c$ and $\Delta_i$. The region covered by the modulation is
marked by the rectangle. Single-tone modulation proceeds along the two lines
crossing the magenta point: horizontally for $\Delta_c$ modulation and vertically
for $\Delta_i$ modulation. Panel (a) shows the variation of $g^{(2)}(t)$ over one
period of the long-time (Floquet) state for each case, in red and blue
respectively. Neither single-tone modulation reaches the antibunching regime. With bi-tone modulation, by contrast, the correlation function drops well below
unity, reaching a minimum of $0.30$. Because the two detunings are modulated at
different frequencies, the system traces a Lissajous trajectory in the
$(\Delta_c,\Delta_i)$ plane, covering a two-dimensional region rather than a
line. This result can be summarised as:
\begin{equation}
\min \,g^{(2)}(t)\Big|_{\text{bi-tone}}
\;<\;
\min_{(\Delta_c,\Delta_i)\in\mathcal{R}}\,g^{(2)}(0)\Big|_{\Lambda,\phi\ \text{fixed}}\,,
\label{eq:dynbound}
\end{equation}
where $\mathcal{R}=[\Delta_c^0-A_c,\Delta_c^0+A_c]\times[\Delta_i^0-A_i,\Delta_i^0+A_i]$
denotes the region of the detuning plane swept by the modulation. The effect is not, however, simply a matter of sampling favourable static
configurations: the lowest value of $g^{(2)}(0)$ anywhere within the rectangle is
$0.46$, and the deepest dynamical antibunching occurs at detunings where the
static correlation function is bunched. The modulation therefore establishes an
interference condition that is absent from the static phase diagram and
inaccessible to single-tone driving.

In summary, the dynamical photon statistics in our Raman-coupled cavity QED system exhibit rich and controllable behavior under time modulation of the detuning. Single-tone modulation can enhance the antibunching when the static system is already in the blockade regime, and can partially restore antibunching even when the static parameters are not optimal, although full restoration requires careful parameter selection. A direct consequence of the time-dependent Raman interaction (mediated by the modulated detuning) is the emergence of controlled emission windows, providing a route toward time-gated single-photon generation compatible with standard Raman configurations \cite{Mundhada2019}. Such temporal control may be relevant for quantum communication protocols involving time-bin encoded photons \cite{Ritter2025} and for synchronizing photon emission with external clock signals in quantum networks. The demonstrated ability to dynamically sculpt photon statistics through time modulation in the Raman coupling system opens up new possibilities for quantum information processing and quantum communication. By choosing the modulation parameters appropriately, one can generate tailored temporal profiles of $g^{(2)}(t)$ that are optimized for specific applications---for example, creating long windows of antibunching for quantum memory operations, or rapid switching between bunching and antibunching for time-division multiplexing in quantum networks. The flexibility of our scheme, combined with the availability of experimental techniques for modulating detunings in cavity QED systems, makes it a promising platform for dynamic quantum photonics.

\section{Conclusion}
We studied photon statistics through quantum interference in a  Raman-coupled cavity-atom system with parametric amplification. Our analytical and numerical results reveal a distinct role of Raman coupling in controlling interference-induced photon correlations, which has not been explored in previous platforms
~\cite{spb4-kgmq,PhysRevLett.129.043601,doi:10.1126/sciadv.abj1916,LeJeannic2022,Ghosh2019,PhysRevA.110.023718}. Specifically, we find that the Raman interaction, combined with the parametric drive, provides a tunable control parameter that modifies the photon statistics from antibunching toward bunching in both steady-state and dynamical regimes.
In the steady-state regime, manipulating the Raman coupling disrupts the destructive interference responsible for photon blockade, driving the system from antibunched toward bunched photon correlations. While our analysis focuses on the onset of this crossover, the underlying model shows that larger Raman coupling can further enhance bunching. Nontrivial dynamical behavior emerges when the atomic and cavity frequencies are periodically modulated in time. Under single-tone modulation, we observe a periodic enhancement of antibunching/bunching and the system can dynamically evolve between bunched and antibunched photon statistics. Extending the modulation scheme to bi-tone modulation induces antibunching even in parameter regimes where the steady state is bunched and single-tone modulation does not generate nonclassical correlations. The effect is not merely a matter of sweeping through favourable static configurations: the dynamical minimum lies below the lowest value attainable anywhere within the swept parameter region. A direct consequence of the time modulation is to provide a route toward time-gated single-photon generation compatible with standard Raman configurations~\cite{Mundhada2019}. Such temporal control may be relevant for quantum communication protocols involving time-bin encoded photons~\cite{Ritter2025} and for synchronizing photon emission with external clock signals in quantum networks.
Overall, our results demonstrate that Raman coupling provides a tunable mechanism for controlling interference-induced photon correlations, while temporal modulation enables their dynamical shaping.

\section*{Acknowledgments}
AR  acknowledges funding from the European Union’s Horizon Europe research and innovation programme under grant agreements No. ID 101115575 (Q-ONE) and ID 101130304 (PolArt). The C4QEC project is carried out within the IRAP of the Foundation for Polish Science co-financed by the European Union.


\bibliographystyle{apsrev4-2}

\bibliography{refs}

\newpage
\onecolumngrid
\begin{center}
\textbf{Supplemental Material for Dynamic Sculpting of Photon Statistics in a Raman-Coupled Cavity QED System}
\end{center}
\appendix

\section{ Microscopic Origin of the Raman Coupling and Its Role as a Control Switch}
We consider a 3-level atom with ground states $\vert{}g\rangle, \vert{}e\rangle$  and an excited state $\vert{}i\rangle$ and its coupling to a cavity mode. The Hamiltonin in the lab frame is detailed below, where we assume $\vert{}g\rangle$ as the zero energy reference :

\begin{align}\label{eq:mainHam}
\hat{H}_{\text{lab}}(t) = \hat{H}_{\text{free}} + \hat{H}_{\text{atom-field}}(t) + \hat{H}_{\text{cav-drives}}(t)
\end{align}
 
\begin{align}
\hat{H}_{\text{free}} = \hbar \omega_c \hat{a}^\dagger \hat{a} + \hbar \omega_{eg} \vert{}e\rangle\langle e\vert{} + \hbar \omega_{ig} \vert{}i\rangle\langle i\vert{}
\end{align}

State $\vert{}i\rangle$ couples to both lower states via the quantized cavity mode ($g_1, g_2$) and two classical lasers (Pump $\Omega_p$ at $\omega_{\text{pump}}$, Stokes $\Omega_s$ at $\omega_{\text{stokes}}$):

\begin{align}
\hat{H}_{\text{atom-field}}(t) &= \hbar g_1 \left( \hat{a} \vert{}i\rangle\langle g\vert{} + \hat{a}^\dagger \vert{}g\rangle\langle i\vert{} \right) + \hbar g_2 \left( \hat{a} \vert{}i\rangle\langle e\vert{} + \hat{a}^\dagger \vert{}e\rangle\langle i\vert{} \right)\nonumber\\&+ \frac{\hbar \Omega_p}{2} \left( \vert{}i\rangle\langle g\vert{} e^{-i \omega_{\text{pump}} t} + \vert{}g\rangle\langle i\vert{} e^{i \omega_{\text{pump}} t} \right) + \frac{\hbar \Omega_s}{2} \left( \vert{}i\rangle\langle e\vert{} e^{-i \omega_{\text{stokes}} t} + \vert{}e\rangle\langle i\vert{} e^{i \omega_{\text{stokes}} t} \right)
\end{align}

\begin{align}
\hat{H}_{\text{cav-drives}}(t) = \hbar \Omega \left( \hat{a}^\dagger e^{-i \omega_d t} + \hat{a} e^{i \omega_d t} \right) + \hbar \Lambda \left( \hat{a}^{\dagger 2} e^{-i(2\omega_d t + \phi)} + \hat{a}^2 e^{i(2\omega_d t + \phi)} \right)
\end{align}

In general, finding a rotating frame in which all time-dependencies can be discarded is not possible for the Hamiltonian in Eq.~(\ref{eq:mainHam}). Instead, we focus on a simpler case where the time dependency can be removed under some conditions~\footnote{It is worth noting there are other variant for which the time dependency can be removed, here we are interested to the cavity-mediated Raman problem}. 
To this end, we assume $\Omega_p=\Omega_s=0$ which correspond to cavity-mediated Raman scattering. Using $\hat{U}(t) = \exp\left[-i \omega_d t \left( \hat{a}^\dagger \hat{a} + \vert{}i\rangle\langle i\vert{} \right)\right]$, we obtain the following time-independent Hamiltonian:
\begin{align}\label{eq:A5}
\hat{H} &= \hbar \Delta_c \hat{a}^\dagger \hat{a} + \hbar \omega_{eg} \vert{}e\rangle\langle e\vert{} + \hbar \Delta_i \vert{}i\rangle\langle i\vert{}\nonumber\\ &\quad+ \hbar g_1 \left( \hat{a} \vert{}i\rangle\langle g\vert{} + \hat{a}^\dagger \vert{}g\rangle\langle i\vert{} \right) + \hbar g_2 \left( \hat{a} \vert{}i\rangle\langle e\vert{} + \hat{a}^\dagger \vert{}e\rangle\langle i\vert{} \right) \nonumber\\&\quad+ \hbar \Omega \left( \hat{a}^\dagger + \hat{a} \right) + \hbar \Lambda \left( \hat{a}^{\dagger 2} e^{-i\phi} + \hat{a}^2 e^{i\phi} \right)
\end{align}
where we define the detunings $\Delta_c = \omega_c - \omega_d$ and $\Delta_i = \omega_{ig} - \omega_d$.
%
We define the jump operators: $\hat{L}_g=\sqrt{\gamma_g}\vert{}g\rangle\langle i\vert{}$, $\hat{L}_e=\sqrt{\gamma_e}\vert{}e\rangle\langle i\vert{}$, and $\hat{L}_a=\sqrt{\kappa}\hat{a}$. We are interested in the case that excited state $|i\rangle$ can be eliminated. This leads to an effective Hamiltonian in which the two lower ground states are coupled through Raman interaction. Following the method of Ref.~\cite{PhysRevA.85.032111}, we introduce the Hamiltonian of ground $H_g$ and excited $H_e$ subspaces:
\begin{align}
\hat{H}_g = \hat{\mathbb{I}}_{\text{c}} \otimes \hbar \omega_{eg} \vert{}e\rangle\langle e\vert{} + \hat{H}_{\text{c}} \otimes \left( \vert{}g\rangle\langle g\vert{} + \vert{}e\rangle\langle e\vert{} \right)\,,
\end{align}
\begin{align}
\hat{H}_e = \hat{\mathbb{I}}_{\text{c}} \otimes \hbar \Delta_i \vert{}i\rangle\langle i\vert{} + \hat{H}_{\text{c}} \otimes \vert{}i\rangle\langle i\vert{}\,,
\end{align}
where $\hat{H}_{\text{c}} = \hbar \Delta_c \hat{a}^\dagger \hat{a} + \hbar \Omega \left( \hat{a}^\dagger + \hat{a} \right) + \hbar \Lambda \left( \hat{a}^{\dagger 2} e^{-i\phi} + \hat{a}^2 e^{i\phi} \right)$ and $\hat{\mathbb{I}}_{\text{c}}$ is identity $N \times N$ matrix. We also define (de-)excitation term between the ground and excited subspaces:
\begin{align}
\hat{V}_+ =& \hbar \hat{a}\otimes \vert{}i\rangle \left( g_1 \langle g\vert{} + g_2 \langle e\vert{} \right)\nonumber\\
\hat{V}_- =& \hat{V}_+^\dagger = \hbar \hat{a}^\dagger\otimes \left( g_1 \vert{}g\rangle + g_2 \vert{}e\rangle \right) \langle i\vert{}
\end{align}
We notice that $\hat{H} = \hat{H}_g + \hat{H}_e + \hat{V}_+ + \hat{V}_-$. Given this, we use the following definitions:
\begin{align}
\hat{H}_{\mathrm{eff}}
&= -\frac{1}{2}\hat{V}_{-}
\left(
\hat{H}_{\mathrm{NH}}^{-1}
+
\left(\hat{H}_{\mathrm{NH}}^{-1}\right)^\dagger
\right)
\hat{V}_{+}
+\hat{H}_{g},
\\[6pt]
\hat{L}_{\mathrm{eff}}^{k}
&= \hat{L}_{k}\hat{H}_{\mathrm{NH}}^{-1}\hat{V}_{+}\,,
\end{align}
with $\hat{H}_{\text{NH}}=\hat{H}_e-\frac{i}{2}\sum_{j \in \{g,e,a\}}\hat{L}_j^\dagger \hat{L}_j$. This gives the following results:  
\begin{align}
\hat{H}_{\text{eff}} =\hat{\mathbb{I}}_c\otimes\hbar\omega_{eg}|e\rangle\langle e|+H_\text{c} \otimes \hat{\mathbb{I}}_a - \hat{\mathcal{C}} \otimes \hat{\mathcal{A}}
\end{align}
with $\hat{\mathcal A}=\frac{g_1^2+g_2^2}{2}\hat{\mathbb{I}}_a+
g_1g_2\sigma_x+\frac{g_2^2-g_1^2}{2}\sigma_z$, $\sigma = |e\rangle\langle g|$, and $\hat{\mathcal{C}} = \hat{a}^\dagger \frac{1}{2\hbar}\bigg[ \sum_{n=0}^{\infty}  \frac{\vert{}n\rangle_R {}_L\langle n\vert{}}{e_0 + \delta n} +\frac{\vert{}n\rangle_L {}_R\langle n\vert{}}{e_0^\ast + \delta^\ast n}   \bigg]\hat{a}$.

To simplify the process of calculating $\hat{H}_{\text{NH}}^{-1}$, we introduce biorthogonal dressed Fock bases $\vert{}n\rangle_R$ and ${}_{L}\!\langle n\vert{}$ with ${}_{L}\!\langle m\vert{}n\rangle_{R}=\delta_{mn}$~\cite{Balian1969}. Because the Hamiltonian is non-Hermitian, it is diagonalized by a non-unitary similarity transformation $U$. We introduce the new biorthogonal bosonic operators $b = U a U^{-1}$ and $\bar{b} = U a^\dagger U^{-1}$, which satisfy $[b,\bar b]=1$ with $b^\dagger\neq\bar b$.To find the diagonal form of $H_{\text{NH}}$, we express the original operators in terms of this new basis. Namely,  we make the following transformations: $a = \cosh(r) b - e^{-i\phi} \sinh (r) \bar{b} + \alpha$ and $\bar{a} = \cosh (r) \bar{b} - e^{i\phi} \sinh (r) b + \beta$. We choose the complex free parameters $\alpha,~\beta,~r$ in such a way that $H_{\text{NH}}$ is diagonal in the photonic subspace. Straightforward calculations yield $\alpha = \Omega \frac{2\Lambda e^{-i\phi} - \tilde{\Delta}_c}{\tilde{\Delta}_c^2 - 4\Lambda^2}$, $\beta = \Omega \frac{2\Lambda e^{i\phi} - \tilde{\Delta}_c}{\tilde{\Delta}_c^2 - 4\Lambda^2}$ and $\tanh(2r) = \frac{2\Lambda}{\tilde{\Delta}_c}$. We also define $\delta = \sqrt{\tilde{\Delta}_c^2 - 4\Lambda^2}$, where $\tilde{\Delta}_c = \Delta_c - i\frac{\kappa}{2}$, $\tilde{\Delta}_i = \Delta_i - i\frac{\gamma_g+\gamma_e}{2}$, and $e_0 = \tilde{\Delta}_i+\Omega^2\frac{2\Lambda \cos\phi-\tilde{\Delta}_c}{\tilde{\Delta}_c^2-4\Lambda^2}$.

Similarly we can show the following relations:
\begin{align}
\hat{L}_{\text{eff}}^{\gamma_g} = \sqrt{\gamma_g} \left[ \sum_{n=0}^\infty \frac{1}{e_0 + \delta  n} \vert{}n\rangle_R{}_{L}\!\langle n\vert{} \right] \hat{a} \otimes \Big( g_1 \vert{}g\rangle\langle g\vert{} + g_2 \vert{}g\rangle\langle e\vert{} \Big)\,,
\end{align}
\begin{align}
\hat{L}_{\text{eff}}^{\gamma_e} = \sqrt{\gamma_e} \left[ \sum_{n=0}^\infty \frac{1}{e_0 + \delta n} \vert{}n\rangle_R{}_{L}\!\langle n\vert{} \right] \hat{a} \otimes \Big( g_1 \vert{}e\rangle\langle g\vert{} + g_2 \vert{}e\rangle\langle e\vert{} \Big)\,,
\end{align}

In the limit that $\Delta_i \gg \{\Delta_c, \Omega, \Lambda, \gamma_{g,e}, \kappa\}$, $g_{1} \sqrt{\langle n \rangle} \ll \vert{}\Delta_i\vert{}$, $g_{2} \sqrt{\langle n \rangle} \ll \vert{}\Delta_i\vert{}$ and assuming $g_1=g_2$ and $\omega_{eg}=0$, we have ($\hbar=1$)
\begin{align}
\hat{H}_{\text{eff}} \approx (\Delta_c+\chi) \hat{n} + \Lambda\left(e^{-i\phi}\hat{a}^{\dagger 2} + e^{i\phi}\hat{a}^2\right) + \Omega(\hat{a}^\dagger + \hat{a}) +\chi \hat{n} \otimes \hat{\sigma}_x
\end{align}
with $\chi = -g_0^2 / \Delta_i$.

We now use Eq.~\ref{eq:A5} to approximately find the condition for $g^{(2)}(0)=0$. This solution is valid in the regime of weak drives.  We define $\hat{H}_{\text{Non-Herm}}=\hat{H}-\frac{i}{2}\sum_{j \in \{g,e,a\}}\hat{L}_j^\dagger \hat{L}_j$ 

Let's introduce $\vert{}\psi(t)\rangle = \sum_{n=0}^2 \sum_{\alpha \in \{g,e,i\}} c_{n,\alpha}(t) \vert{}n, \alpha\rangle$ and use $i \dot{\psi} = H_{\text{Non-Herm}} \psi$ to find the equations for $c_{n,\alpha}$ coefficients.
After some algebra we can find the following equations:
\begin{subequations}
\begin{align}
i\dot{c}_{0,g} &= \Omega c_{1,g} + \sqrt{2}\Lambda e^{i\phi} c_{2,g}\\
i\dot{c}_{1,g} &= \tilde{\Delta}_c c_{1,g} + \Omega c_{0,g} + \sqrt{2}\Omega c_{2,g} + g_1 c_{0,i}\\
i\dot{c}_{2,g} &= 2\tilde{\Delta}_c c_{2,g} + \sqrt{2}\Omega c_{1,g} + \sqrt{2}\Lambda e^{-i\phi} c_{0,g} + \sqrt{2}g_1 c_{1,i}\\
i\dot{c}_{0,e} &= \omega_{eg} c_{0,e} + \Omega c_{1,e} + \sqrt{2}\Lambda e^{i\phi} c_{2,e}\\
i\dot{c}_{1,e} &= (\tilde{\Delta}_c + \omega_{eg}) c_{1,e} + \Omega c_{0,e} + \sqrt{2}\Omega c_{2,e} + g_2 c_{0,i}\\
i\dot{c}_{2,e} &= (2\tilde{\Delta}_c + \omega_{eg}) c_{2,e} + \sqrt{2}\Omega c_{1,e} + \sqrt{2}\Lambda e^{-i\phi} c_{0,e} + \sqrt{2}g_2 c_{1,i}\\
i\dot{c}_{0,i} &= \tilde{\Delta}_i c_{0,i} + \Omega c_{1,i} + \sqrt{2}\Lambda e^{i\phi} c_{2,i} + g_1 c_{1,g} + g_2 c_{1,e}\\
i\dot{c}_{1,i} &= (\tilde{\Delta}_c + \tilde{\Delta}_i) c_{1,i} + \Omega c_{0,i} + \sqrt{2}\Omega c_{2,i} + \sqrt{2}g_1 c_{2,g} + \sqrt{2}g_2 c_{2,e}\\
i\dot{c}_{2,i} &= (2\tilde{\Delta}_c + \tilde{\Delta}_i) c_{2,i} + \sqrt{2}\Omega c_{1,i} + \sqrt{2}\Lambda e^{-i\phi} c_{0,i}
\end{align}
\end{subequations}
For the steady state solutions we assume $c_{g0}=1$. We also assume $\Delta_i$ is much larger than the other parameters. These yield:
\begin{align}
(\tilde{\Delta}_c - S_g) c_{1,g} + \sqrt{2}\Omega c_{2,g} - g_R c_{1,e} = -\Omega\\
\sqrt{2}\Omega c_{1,g} + (2\tilde{\Delta}_c - 2S_g) c_{2,g} - 2g_R c_{2,e} = -\sqrt{2}\Lambda e^{-i\phi}\\
\omega_{eg} c_{0,e} + \Omega c_{1,e} + \sqrt{2}\Lambda e^{i\phi} c_{2,e} = 0\\
\Omega c_{0,e} + (\tilde{\Delta}_c + \omega_{eg} - S_e) c_{1,e} + \sqrt{2}\Omega c_{2,e} - g_R c_{1,g} = 0\\
\sqrt{2}\Lambda e^{-i\phi} c_{0,e} + \sqrt{2}\Omega c_{1,e} + (2\tilde{\Delta}_c + \omega_{eg} - 2S_e) c_{2,e} - 2g_R c_{2,g} = 0\,,
\end{align}
where $S_g = \frac{g_1^2}{\tilde{\Delta}_i}, \quad S_e = \frac{g_2^2}{\tilde{\Delta}_i}, \quad g_R = \frac{g_1 g_2}{\tilde{\Delta}_i}$, $S_g = \frac{g_1^2}{\tilde{\Delta}_i}, \quad S_e = \frac{g_2^2}{\tilde{\Delta}_i}, \quad g_R = \frac{g_1 g_2}{\tilde{\Delta}_i}$. We want the condition for which $g^{(2)}(0) = \frac{2 \left( \vert{}c_{2,g}\vert{}^2  \right)}{\left( \vert{}c_{1,g}\vert{}^2 + \vert{}c_{1,e}\vert{}^2 \right)^2}$, is zero. We obtain the following condition:
\begin{align}
\Lambda e^{-i\phi} = \frac{\Omega^2 \left[ \Omega^2 - \omega_{eg}(\tilde{\Delta}_c + \omega_{eg} - S_e) \right]}{(\tilde{\Delta}_c - S_g)\left[ \Omega^2 - \omega_{eg}(\tilde{\Delta}_c + \omega_{eg} - S_e) \right] + g_R^2 \omega_{eg}}
\end{align}
%
%
\section{Study of $g^{(2)}(\tau)$}
We use the Quantum Regression Theorem (QRT) to compute $g^{(2)}(\tau)$. We start from $G^{(2)}(\tau) = \langle \hat{a}^\dagger(t) \hat{a}^\dagger(t+\tau) \hat{a}(t+\tau) \hat{a}(t) \rangle_{\text{ss}}$, where "$\text{ss}$" denotes the steady state ($t \to \infty$). Assuming steady state $\hat\rho_\text{ss}$ and defining $\hat{\rho}_c(0) = \hat{a} \hat{\rho}_{\text{ss}} \hat{a}^\dagger$, the density matrix at delay time $\tau$ is $\hat{\rho}_c(\tau) = e^{\mathcal{L}\tau}\hat{\rho}_c(0)$, with $\mathcal{L}$ as the Liouvillian superoperator. This yields $G^{(2)}(\tau) = \text{Tr}\left[ \hat{n} \, \hat{\rho}_c(\tau) \right]$, with $\hat{n} = \hat{a}^\dagger\hat{a}$. 

For the analytical considerations and in the weak-driving limit, we define:
\begin{align}
M =-i \begin{pmatrix}  \tilde{\Delta}_c & \sqrt{2}\Omega & 0 & 0 & 0 & g_1 & 0 & 0 \\ \sqrt{2}\Omega & 2\tilde{\Delta}_c & 0 & 0 & 0 & 0 & \sqrt{2}g_1 & 0 \\ 0 & 0 & \omega_{eg} & \Omega & \sqrt{2}\Lambda e^{i\phi} & 0 & 0 & 0 \\ 0 & 0 & \Omega & \tilde{\Delta}_c + \omega_{eg} & \sqrt{2}\Omega & g_2 & 0 & 0 \\ 0 & 0 & \sqrt{2}\Lambda e^{-i\phi} & \sqrt{2}\Omega & 2\tilde{\Delta}_c + \omega_{eg} & 0 & \sqrt{2}g_2 & 0 \\ g_1 & 0 & 0 & g_2 & 0 & \tilde{\Delta}_i & \Omega & \sqrt{2}\Lambda e^{i\phi} \\ 0 & \sqrt{2}g_1 & 0 & 0 & \sqrt{2}g_2 & \Omega & \tilde{\Delta}_c + \tilde{\Delta}_i & \sqrt{2}\Omega \\ 0 & 0 & 0 & 0 & 0 & \sqrt{2}\Lambda e^{-i\phi} & \sqrt{2}\Omega & 2\tilde{\Delta}_c + \tilde{\Delta}_i \end{pmatrix}\,.
\end{align}
With $\mathbf{c} = \begin{pmatrix} c_{1,g}, & c_{2,g}, & c_{0,e}, & c_{1,e}, & c_{2,e}, & c_{0,i}, & c_{1,i}, & c_{2,i} \end{pmatrix}^T$ and $\mathbf{s} = -i \begin{pmatrix} \Omega, & \sqrt{2}\Lambda e^{-i\phi}, & 0, & 0, & 0, & 0, & 0, & 0 \end{pmatrix}^T$, we aim at using $\frac{d\mathbf{c}(t)}{dt} = \mathbf{M} \mathbf{c}(t) + \mathbf{s}$ to find $g^{(2)}(\tau)$. Setting $\frac{d\mathbf{c}}{dt} = 0$, the unconditioned steady-state amplitude vector is obtained directly by matrix inversion:$$\mathbf{c}_{\text{ss}} = -\mathbf{M}^{-1}\mathbf{s} = \begin{pmatrix} c_{1,g}^{\text{ss}}, & c_{2,g}^{\text{ss}}, & c_{0,e}^{\text{ss}}, & c_{1,e}^{\text{ss}}, & c_{2,e}^{\text{ss}}, & c_{0,i}^{\text{ss}}, & c_{1,i}^{\text{ss}}, & c_{2,i}^{\text{ss}} \end{pmatrix}^T$$Upon detecting a photon at $\tau = 0$, the action of the annihilation operator $\hat{a}$ on the steady state $\vert{}\psi_{\text{ss}}\rangle \approx \vert{}0,g\rangle + \sum_{n,\alpha} c_{n,\alpha}^{\text{ss}}\vert{}n,\alpha\rangle$ reduces photon numbers according to $\hat{a}\vert{}n\rangle = \sqrt{n}\vert{}n-1\rangle$. The conditional state amplitude vector immediately following the photon jump is:$$\mathbf{C}(0) = \begin{pmatrix} C_{1,g}(0) \\ C_{2,g}(0) \\ C_{0,e}(0) \\ C_{1,e}(0) \\ C_{2,e}(0) \\ C_{0,i}(0) \\ C_{1,i}(0) \\ C_{2,i}(0) \end{pmatrix} = \begin{pmatrix} \sqrt{2}c_{2,g}^{\text{ss}} \\ 0 \\ c_{1,e}^{\text{ss}} \\ \sqrt{2}c_{2,e}^{\text{ss}} \\ 0 \\ c_{1,i}^{\text{ss}} \\ \sqrt{2}c_{2,i}^{\text{ss}} \\ 0 \end{pmatrix}$$with the ground-state component set to $C_{0,g}(0) = c_{1,g}^{\text{ss}}$.For delay times $\tau > 0$, the ground-state amplitude remains effectively fixed at $C_{0,g}(\tau) \approx c_{1,g}^{\text{ss}}$ in the weak-driving regime. This rescales the effective drive on the excited manifolds to $\mathbf{s}_{\text{cond}} = c_{1,g}^{\text{ss}}\mathbf{s}$. The differential equation governing the conditional dynamics is:$$\frac{d\mathbf{C}(\tau)}{d\tau} = \mathbf{M}\mathbf{C}(\tau) + c_{1,g}^{\text{ss}}\mathbf{s}$$Integrating this system yields the formal time-dependent solution using the matrix exponential:$$\mathbf{C}(\tau) = \mathbf{C}_{\text{ss}} + e^{\mathbf{M}\tau} \left[ \mathbf{C}(0) - \mathbf{C}_{\text{ss}} \right]$$where $\mathbf{C}_{\text{ss}} = c_{1,g}^{\text{ss}}\mathbf{c}_{\text{ss}}$ is the asymptotic conditional steady state as $\tau \to \infty$.The conditional photon expectation value $I(\tau) = \langle \hat{a}^\dagger \hat{a} \rangle_\tau$ is evaluated as:$$I(\tau) = \vert{}C_{1,g}(\tau)\vert{}^2 + \vert{}C_{1,e}(\tau)\vert{}^2 + \vert{}C_{1,i}(\tau)\vert{}^2 + 2\left( \vert{}C_{2,g}(\tau)\vert{}^2 + \vert{}C_{2,e}(\tau)\vert{}^2 + \vert{}C_{2,i}(\tau)\vert{}^2 \right)$$And the unconditioned steady-state photon intensity is given by:$$I_{\text{ss}} = \vert{}c_{1,g}^{\text{ss}}\vert{}^2 + \vert{}c_{1,e}^{\text{ss}}\vert{}^2 + \vert{}c_{1,i}^{\text{ss}}\vert{}^2 + 2\left( \vert{}c_{2,g}^{\text{ss}}\vert{}^2 + \vert{}c_{2,e}^{\text{ss}}\vert{}^2 + \vert{}c_{2,i}^{\text{ss}}\vert{}^2 \right)$$Finally, the normalized second-order coherence function is expressed as:
\begin{align}\label{eq:sdjcn9wedfh}
	g^{(2)}(\tau) = \frac{I(\tau)}{I_{\text{ss}}^2}
\end{align}

Within some approximation, Eq.~(\ref{eq:sdjcn9wedfh}) can be reduced to a simpler form. Keeping the leading terms in the weak-drive regime and assuming a large $\Delta_i$, we obtain:

\begin{align}
g^{(2)}(\tau)\approx\left|1+\left(\frac{\sqrt{2}\,c_{2,g}}
{(c_{1,g})^2}-1\right)e^{-i\mathcal{E}\tau}\right|^2\,,
\label{eq:g2closeddsdc}
\end{align}
where we introduce $\mathcal{E}\equiv\tilde{\Delta}_c-\frac{g_1^2}{\tilde{\Delta}_i}
-\frac{g_1^2g_2^2/\tilde{\Delta}_i^{\,2}}
{\tilde{\Delta}_c+\omega_{eg}-g_2^2/\tilde{\Delta}_i-\Omega^2/\omega_{eg}}$. 
\begin{figure}[t]
	\centering
	\includegraphics[width=0.5\linewidth]{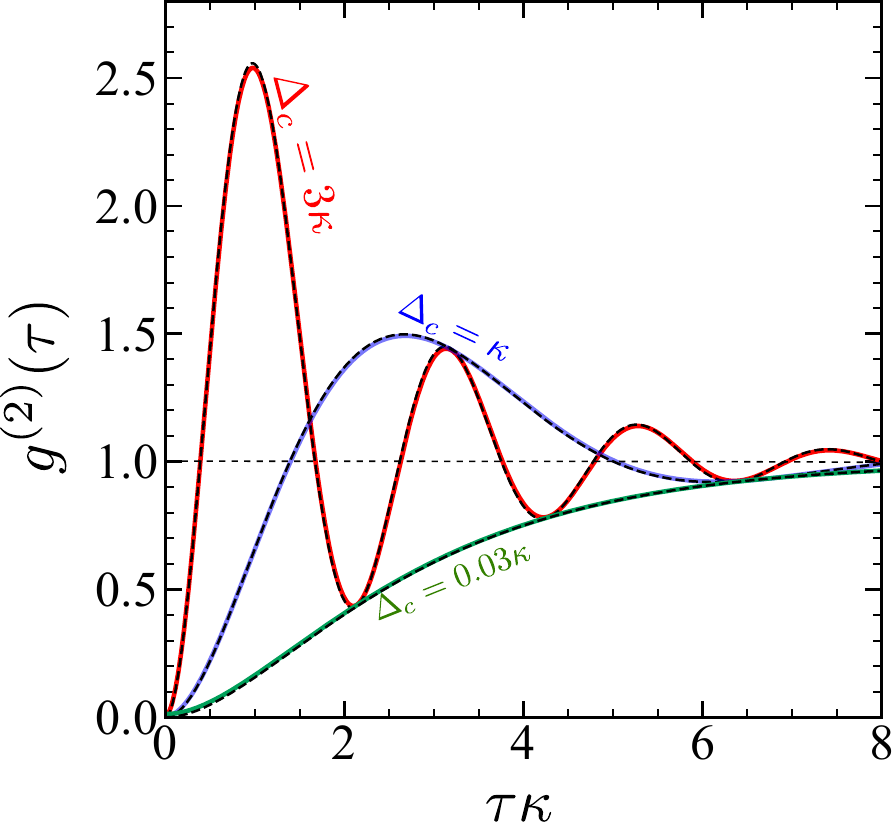}
	\caption{ Second-order correlation function in terms of delay $\tau$. Solid lines shows numerical solution of the master equation. Dashed lines shows the analytical result based on Eq.~(\ref{eq:g2closeddsdc}) Using parameters: $
		g_1=g_2=1\kappa,\quad
		\Omega=0.05\kappa,\quad
		\omega_{eg}=3\kappa,\quad
		\gamma_g=\gamma_e=0.4\kappa
		$, $\Delta_i=20\kappa$. }
	\label{fig:S1}
\end{figure}
Writing $\mathcal{E}=\mathcal{E}'-i\mathcal{E}''/2$, Eq.~(\ref{eq:g2closeddsdc})
becomes
\begin{align}
g^{(2)}(\tau)=1+r^2e^{-\mathcal{E}''\tau}
+2r\,e^{-\mathcal{E}''\tau/2}\cos\!\left(\mathcal{E}'\tau-\theta\right),\label{eq:b4}
\end{align}
where $r e^{i\theta}=\sqrt{2}c_{2,g}/(c_{1,g})^2-1$. The
antibunching therefore recovers through an oscillation at $\mathcal{E}'$, while
the return to $g^{(2)}=1$ proceeds at the rate $\mathcal{E}''$. We may simplify as (with $g_1=g_2$): 
%
\begin{align}
\mathcal{E}'&\simeq\Delta_c-\frac{g_1^2\Delta_i}{\Delta_i^2+\bar\gamma^2},\\
\mathcal{E}''&\simeq\kappa+\frac{2g_1^2\bar\gamma}{\Delta_i^2+\bar\gamma^2},
\end{align}
with $\bar\gamma=(\gamma_g+\gamma_e)/2$. Figure~\ref{fig:S1} compares Eq.~(\ref{eq:g2closeddsdc}) with the numerical
solution of the master equation for three values of the cavity detuning, showing good agreement. We can provide an analytical expression for the $\tau$ window for which the light remains antibunched. At the blockade condition 
Eq.~(\ref{eq:b4}) gives $g^{(2)}(\tau)=1+e^{-\mathcal{E}''\tau}
-2e^{-\mathcal{E}''\tau/2}\cos(\mathcal{E}'\tau)$, which first returns to unity
when $e^{-\mathcal{E}''\tau/2}=2\cos(\mathcal{E}'\tau)$. For
$\mathcal{E}'\gg\mathcal{E}''$ this yields $\tau_{\rm w}\simeq\pi/3\mathcal{E}'$:
the delay window over which the light remains antibunched is set by the
effective detuning rather than by the cavity lifetime. 
%

\end{document}